\documentclass{aa}

\usepackage{graphicx}
\usepackage{txfonts}
\usepackage{caption}
\usepackage{subcaption}
\usepackage{amsmath,amsfonts,bm}
\usepackage{comment}
\usepackage{orcidlink}
\usepackage{cleveref}

\usepackage{xcolor}

\newcommand{\legolas}{\textsf{Legolas}}
\newcommand{\amrvac}{\textsf{MPI-AMRVAC}}

\begin{document} 

   \title{Streamer slab eigenmode analysis with the \legolas{} code}

   \author{J. De Jonghe\inst{1} \orcidlink{0000-0003-2443-3903}, D. Sorokina \inst{1} \orcidlink{0000-0003-3351-4326} and T. Van Doorsselaere\inst{1} \orcidlink{0000-0001-9628-4113}}

   \institute{Centre for mathematical Plasma Astrophysics, Department of Mathematics, 
             KU Leuven, Celestijnenlaan 200B, 3001 Leuven, Belgium. 
             \email{jordi.dejonghe@kuleuven.be}
             }

\date{\today}
\titlerunning{Streamer slab eigenmode analysis with the \legolas{} code}
\authorrunning{De Jonghe et al.}
 
    \abstract
   {Helmet streamers are large ray-like structures extending from the solar corona that thin further away from the solar surface, forming an extended current sheet. These structures are quasi-stable, and are observed to support kink waves travelling outward from the Sun, called helmet streamer waves.}
   {Limited analytical models identify streamer waves as fast body kink eigenmodes of the streamer slab. To bridge the gap between analytical models and numerical simulations, we investigate the eigenmode spectrum of more realistic streamer slab configurations, obtained from simulations, to retrieve the kink profiles and firmly establish that streamer waves are eigenmodes of the streamer.}
   {Using a streamer slab model extracted from numerical simulations, the \legolas{} code is applied to compute the eigenmode spectrum, for wavelengths observed in the simulation. In the spectrum we identify the mode matching the behaviour in the non-linear simulations by comparing to analytical and numerical results for the established Epstein profile. The identified modes's phase speeds are then compared to analytical, simulation, and observational results.}
   {For each case, the spectrum is found to contain a mode matching the properties of a streamer wave. The phase speeds of the identified modes are compatible with those measured in observations, strongly suggesting that the observed streamer waves are fast body kink modes. Comparison to the analytical slab model is difficult due to the sensitivity of the analytical dispersion relation to the internal values.}
   {}

   \keywords{}

   \maketitle
   \nolinenumbers

\section{Introduction} \label{sec:introduction}

{In this work, we discuss the eigenvalue problem for helmet streamers —large, ray-like coronal structures shaped like arches near the Sun that narrow to thin stalks ($\sim$ 3 $^{\circ}$ wide at 5 $R_\odot$) and can extend up to 30 $R_\odot$ \citep{koutchmy92,loucif92}. Helmet streamers are quasi-stable (covering several solar rotations) structures formed above active regions, where closed magnetic flux connects opposite polarities. A cusp forms above the closed field, extending as a current sheet to the heliospheric current sheet between open flux systems \citep{koutchmy71,Zhukov08}.

Of particular interest are the largest wave phenomena in the solar corona: streamer waves. These are\ transverse oscillations propagating outwards along the plasma sheet of the thin streamer stalk and decaying within a few periods or wavelengths \citep{Chen10,Feng11}. They are generated by the solar coronal dynamics; for example,\ a streamer wave can be excited by disturbances due to interaction with a coronal mass ejection (CME).

Such waves are valuable candidates for coronal seismology \citep[see][]{DN,NK}, a technique applied by \citet {Chen10} and \citet {Feng11} to infer Alfvén speed profiles and magnetic field strengths by treating the waves as fast kink body modes \citep{edwin} and linking their observed phase speeds to the external Alfv\'en speed. A more in-depth analysis of the dispersion relation for a streamer-slab-like configuration, where a non-uniform plasma slab and an alternating magnetic field were considered, was presented in Chapter~4 of \citet {thesis}. Consistently with similar numerical studies by \citet {smith} and \citet {fruit}, \citet {thesis} showed that the phase speed of the kink oscillation is approximately equal to the external Alfv\'en speed.

\citet {Decraemer20} also proposed that seismology can be used to precisely determine the solar wind speed because of the Doppler shift in magnetohydrodynamics (MHD) wave theory \citep[e.g.][]{goossens03,N}. In \citet {Chen10}, \citet {Feng11} and \citet {Decraemer20} the measured speeds were found to be significantly higher than the calculated speeds (proper phase speed of the wave) because of the Doppler shift, i.e.\ the contribution from the background solar wind. By combining the streamer wave speed obtained from the observations with the proper phase speed of the wave in the framework of well-developed streamer wave theory, we can estimate the solar wind speed. This seismology technique is an alternative to the currently used blob tracking \citep{Sheeley,Wang}, which is limited by the required assumption of exact equality between the blob speed and the solar wind speed.

A step forward to expand our knowledge of streamer waves through numerical modelling was taken in \citet{Sorokina24}. \citet{Sorokina24} constructed a model of a helmet streamer and observed the oscillation of the streamer stalk triggered by a velocity perturbation. Exploiting this model, the authors performed a parameter study to investigate the sensitivity of the streamer dynamics to the background solar wind speed, and the input parameters for the model, such as temperature and magnetic field, and complemented the statistical analysis of these events comparing numerical and observational findings.

In this paper, we bridge the gap between analytical streamer slab models and numerical streamer simulations, establishing that streamer waves are fast body kink eigenmodes of the system using numerical MHD spectroscopy. To do so, we revisit the \amrvac{} \citep{Keppens2023} simulations presented in \citet{Sorokina24}. Using the numerical code \legolas{} \citep{claes20,Claes23}, which allows coupling with \amrvac{} in its current implementation, we perform an MHD spectroscopic analysis of the simulation data to find the system's natural oscillations with streamer wave properties.

\section{MHD waves in presence of a background flow} \label{sec:equations}

In the solar corona waves propagate in an already moving medium, the solar wind. Since background flow modifies the spectral properties of waves, we briefly review Doppler shift in MHD wave theory here.

For advected quantities such as Eulerian entropy perturbation, the operator $\partial_t+\pmb{v}_0\cdot \nabla$ with time $t$ and equilibrium velocity $\pmb{v}_0$ can vanish, leading to the entropy continuum. It appears at a Doppler-shifted frequency, but does not correspond to genuine plasma oscillations and is typically mathematical. For a more detailed explanation, we refer to \citet{Goedbloed19}. To derive the dispersion relation in presence of a flow one may thus consider (see, e.g., \citealt{goossens92,Goedbloed19}) the Lagrangian representation in terms of the three components of a displacement vector $\pmb{\xi}$ instead of the Eulerian perturbations. The physically significant Lagrangian continua (Alfv\'en and slow) are also shifted by the background flow. They correspond to the vanishing of the same convective derivative acting on the variables that are expressible in terms of $\pmb{\xi}$ and represent real oscillations in the plasma that propagate with the flow, now observed at Doppler-shifted frequencies.

In this section we will not consider in detail the Lagrangian representation and refer to \citet{goossens92,Goedbloed19} instead. Similar to \citet{goossens92} we introduce the Doppler-shift relation
\begin{equation}
\Omega=\omega-\omega_f
\end{equation}
between the rest frame (co-moving) frequency $\Omega$ and the lab frame frequency $\omega$, where $\omega_f$ is the equilibrium flow frequency
\begin{equation}
\omega_f=kv_0(x).
\end{equation}
Here, $v_0$ is the equilibrium flow speed as a function of position, assumed to be in the $z$-direction parallel to the magnetic field $\pmb{B}_0$, and $k=k_z$ (in notations of \citealt{goossens92}) is the wavenumber in that direction.

\subsection{Waves on a magnetic slab}\label{sec:theory}
Let us first consider a uniform plasma inside and outside the slab, separated by two discontinuous interfaces across which the perturbations are continuous, and neglect gravity, as in \citet{edwin}. For a slab of width $2a$, we distinguish between internal (subscript $i$) and external (subscript $e$) values of the plasma parameters,
\begin{equation}\label{eq:anal-init}
  B_0(x), \rho_0(x), p_0(x), v_0(x) =
    \begin{cases}
      B_e, \rho_e, p_e, v_e & \text{for $|x| > a$},\\
      B_i, \rho_i, p_i, v_i & \text{for $|x| < a$},
    \end{cases}       
\end{equation}
where $B$, $\rho$, $p$, and $v$ represent magnetic field strength, density, thermal pressure, and flow speed, respectively.

For this model, \citet{edwin} showed that the dispersion relation for kink waves, including both internal (subscript $i$) and external (subscript $e$) steady flows, is given by
\begin{equation}
 \rho_i (\Omega_i^2-\omega_{Ai}^2) m_e + \rho_e (\Omega_e^2-\omega_{Ae}^2) m_i \text{coth}(m_ia)=0,
\end{equation}
or, in expanded form, dividing by $k^2$,
\begin{equation}\label{eq:full-DR}
 \rho_i [(v_{ph}-v_i)^2-v_{Ai}^2] m_e + \rho_e [(v_{ph}-v_e)^2-v_{Ae}^2] m_i \text{coth}(m_ia)=0,
\end{equation}
where $\omega_A$ and $v_A = B/\sqrt{4\pi\rho}$ denote the Alfv\'en frequency and speed, and $v_{ph}$ the phase speed in the lab frame. Using $\Omega=k(v_{ph}-v_e)$ and introducing the adiabatic coefficient $\gamma$, $m_e$ may be written as
\begin{align}
m_e^2&=\frac{(k^2 c_{se}^2-\Omega_e^2)(k^2v_{Ae}^2-\Omega_e^2)}{(c_{se}^2+v_{Ae}^2)(k^2c_{Te}^2-\Omega_e^2)}\\&=k^2\frac{[c_{se}^2-(v_{ph}-v_e)^2][v_{Ae}^2-(v_{ph}-v_e)^2]}{(c_{se}^2+v_{Ae}^2)[c_{Te}^2-(v_{ph}-v_e)^2]}
\end{align}
for sound $(c_s)$ and cusp $(c_T)$ speeds
\begin{equation}
    c_s^2 = \frac{\gamma p_0}{\rho_0}, \quad c_T^2 = \frac{c_s^2 v_A^2}{c_s^2 + v_A^2}.
\end{equation}
Analogously, $m_i^2$ is defined using the internal values.

If we now consider a slender slab approximation $ka \ll 1$, the hyperbolic cotangent may be approximated as $\coth(m_ia) = (m_ia)^{-1}$ since $m_i \propto k$. The dispersion relation then reduces to
\begin{equation}\label{eq:kink-disprel}
 \rho_e ((kv_{ph}-kv_e)^2-k^2v^2_{Ae}) + \rho_i ((kv_{ph}-kv_i)^2-k^2v^2_{Ai}) m_e a=0.
\end{equation}
Organised by powers of $ka$, Eq. \ref{eq:kink-disprel} can be rewritten as
\begin{align}\label{eq:power-k}
& (ka)^2\rho_e [(v_{ph}-v_e)^2-v^2_{Ae}] + \notag \\
& (ka)^3\rho_i \left[((v_{ph}-v_i)^2-v^2_{Ai}) \sqrt{\frac{(c_{se}^2-v^2_{ph})(v_{Ae}^2-v^2_{ph})}{(c_{se}^2+v_{Ae}^2)(c_{Te}^2-v^2_{ph})}}\right]=0.
\end{align}
Due to the slender slab approximation $ka\ll1$, we may neglect the second term in Eq. \ref{eq:power-k} and show that
\begin{equation}
(v_{ph}-v_e)^2-v^2_{Ae}=0,
\end{equation}
yielding two solutions,
\begin{equation}\label{eq:anal-expected}
v^{\pm}_{ph}=v_e\pm v_{Ae}.
\end{equation}
The sign $\pm$ corresponds to a wave propagation direction, respectively in the positive and negative $z$-direction, i.e.\ down- and upstream. 

We may distinguish the two cases. If the waves propagate in a sub-Alfv\'enic flow $v_e<v_{Ae}$, $v^{-}_{ph}<0$ corresponds to a backward propagating wave and $v^{+}_{ph}>0$ corresponds to a forward propagating wave. If the waves propagate in a super-Alfv\'enic flow $v_e>v_{Ae}$, both solutions become forward propagating ($v^\pm_{ph}>0$). Nevertheless, to distinguish them we will call them the forward and backward propagating solutions regardless of the regime.

In the solar corona streamer waves may propagate in both sub-Alfv\'enic and super-Alfv\'enic flows. Thus, theoretically streamer waves can be backward propagating. However, observers have identified them as forward propagating \citep{Chen10, Feng11, Decraemer20}. The possible reason for this is the damping due to the resonant absorption when the spatial damping depends on direction of propagation \citep{goossens92,goossens11}. Thus, there is an increased damping of the backward propagating waves.

\subsection{Epstein profile}

For a smoothly inhomogeneous plasma, analytical results were obtained for the Epstein profile by \citet{Nakariakov95} in low-$\beta$ plasmas. There they assume a uniform magnetic field $\pmb{B} = B_0\pmb{e}_z$ and density profile
\begin{equation}\label{eq:epstein}
    \rho(x) = \rho_e + (\rho_i-\rho_e)\,\mathrm{sech}^2\left(\frac{|x|}{a}\right)
\end{equation}
such that $\rho = \rho_i$ at $x=0$ and smoothly transitions to $\rho = \rho_e$ as $|x|\rightarrow\infty$. Under these assumptions, the dispersion relation of the kink wave becomes
\begin{equation}\label{eq:epstein-dispersion}
    v_{Ai}^2 \sqrt{v_{Ae}^2 - v_{ph}^2} = v_{Ae}|k|a(v_{ph}^2-v_{Ai}^2),
\end{equation}
where $v_{Ai}$ and $v_{Ae}$ are the Alfv\'en speeds for densities $\rho_i$ and $\rho_e$, respectively, and $v_{Ai} < v_{ph} < v_{Ae}$. Furthermore, the kink mode's transverse velocity perturbation $v_x = U(x)\exp(\mathrm{i}\omega t - \mathrm{i}kz)$ is
\begin{equation}\label{eq:epstein-U}
    U(x) = \mathrm{sech}^\nu\left(\frac{x}{a}\right), \quad \nu = \frac{|k|a}{v_{Ae}} \sqrt{v_{Ae}^2-v_{ph}^2}.
\end{equation}
We validate the use of the \legolas{} code against this case in Sec. \ref{sec:legolas-epstein}.

\section{Numerical methods} \label{sec:setup}

After reaching the steady-state from \citet{Sorokina24}, an eigenmode analysis of the simulations is performed with the numerical code \legolas{}, to identify the streamer slab's eigenmodes with streamer wave properties. In this section we briefly discuss the basics of the \legolas{} code. For a detailed discussion, we refer to the code papers \citep{claes20,Claes23} or the website (\url{https://legolas.science}). Though the code features a modular implementation of a wide range of physical effects, such as background flow, gravity, resistivity, energy losses/gains, viscosity, and Hall MHD \citep{Jordi22}, we limit ourselves to ideal MHD with background flow here.

For any three-dimensional (3D) Cartesian or cylindrical MHD equilibrium with non-trivial 1D variation, \legolas{} quantifies the system's natural oscillations. Concretely, this means that the equilibrium quantities, indicated with a subscript $0$, depend only on a single coordinate, chosen to be $x$ in Cartesian geometry, with no vector components along this coordinate direction. Hence, the configuration is represented by a density $\rho_0=\rho_0(x)$, temperature $T_0=T_0(x)$, velocity $\pmb{v}_0=v_{02}(x)\,\pmb{e}_2+v_{03}(x)\,\pmb{e}_3$ and magnetic field $\pmb{B}_0=B_{02}(x)\,\pmb{e}_2+B_{03}(x)\,\pmb{e}_3$, where $\pmb{e}_2$ and $\pmb{e}_3$ are the unit vector in the $y$- and $z$-direction, respectively.

After perturbing density, temperature, velocity and magnetic field, Fourier analysis is applied to the linearised equations, assuming all perturbations $f_1 \in \{\rho_1, T_1, \pmb{v_1}, \pmb{B_1}\}$ take the form
\begin{equation}\label{eq:fourier}
    f_1(\pmb{r},t)=\hat{f}_1(x)\exp[\mathrm{i}(k_2y+k_3z-\omega t)],
\end{equation}
with wave vector $\pmb{k}=k_2\,\pmb{e}_y+k_3\,\pmb{e}_z$ as input, and angular frequencies $\omega$ and amplitudes $\hat{f}_1(x)$ as output. A detailed derivation of the linearised equations can be found in \citet{Niels_thesis}. With a finite element discretisation, the problem is reduced to an eigenvalue problem \citep[see][]{Goedbloed19,claes20}, which is solved to obtain the couples of complex eigenvalues $\omega$ and state vectors $\pmb{\text{x}}=(\rho_1, \pmb{v_1}, T_1, \pmb{B_1})^T$ corresponding to the fundamental linear waves of the system. The oscillation frequency is represented by $\mathrm{Re}(\omega)$, and $\mathrm{Im}(\omega)$ quantifies the growth ($\mathrm{Im}(\omega) > 0$) or damping rate ($\mathrm{Im}(\omega) < 0$). Using Eq. \ref{eq:fourier}, the system's eigenmodes can be visualised in multiple dimensions by specifying the temporal and spatial coordinates ($t,y,z$) to generate temporal evolutions and spatial views.

Since \legolas{} assumes a 1D spatial variation, unlike the 2D \amrvac{} simulations presented in \citet{Sorokina24}, we consider one-dimensional slits at fixed radial distance $r_\mathrm{slit}$, extracting field data along those slits from a selection of these simulations. Neglecting the slit curvature, the gravitational stratification perpendicular to the slit, and the small vector components along the slit, this results in a Cartesian approximation of the numerical streamer slab model to analyse with \legolas{}. In Fig.~\ref{fig:simulation} we show a simulation snapshot with slits indicated at various distances $r_\mathrm{slit}$. The variation of the `equilibrium' quantities is along the angular $\theta$-direction and the wave propagation is in the simulation’s radial direction. Hence, for our \legolas{} analysis we consider $s = r_\mathrm{slit}(\theta-\pi/2)$ as the $x$-coordinate in our Cartesian approximation, along which our initial state varies, i.e. $\rho_0(s)$, $T_0(s)$, $\pmb{v}_0=v_{03}(s)\,\pmb{e_3}$, and $\pmb{B}_0=B_{03}(s)\,\pmb{e_3}$. Then $r_{\mathrm{slit}}\phi$ is considered to be the $y$-coordinate and $r$ the $z$-coordinate in this Cartesian approximation. The slab is considered infinite and uniform in the directions perpendicular to the slit (i.e. $y,z$) with $B_{02}=0$, $v_{02}=0$. The wave vector is restricted to be in the $r$-direction, $\pmb{k} = k_3\,\pmb{e}_3$, essentially reducing the analysis to 2D ignoring azimuthal variation ($y$ in the Cartesian approximation). This implies that the waves travel radially outward, neglecting any `azimuthal' ($k_2 = k_\phi$) propagation that may be excited by the non-radial path of the exciting CME. This assumption also neglects the magnetic field curvature due to solar rotation (Parker spiral). Perfectly conducting wall boundaries are then applied in the $s$-direction. To prevent any influence from the boundary conditions, we place the boundaries sufficiently far away \citep[see][and Sec. \ref{sec:legolas-epstein}]{Zaliznyak03, VanDoorsselaere2004}.

\begin{figure}
\includegraphics[width=0.5\textwidth]{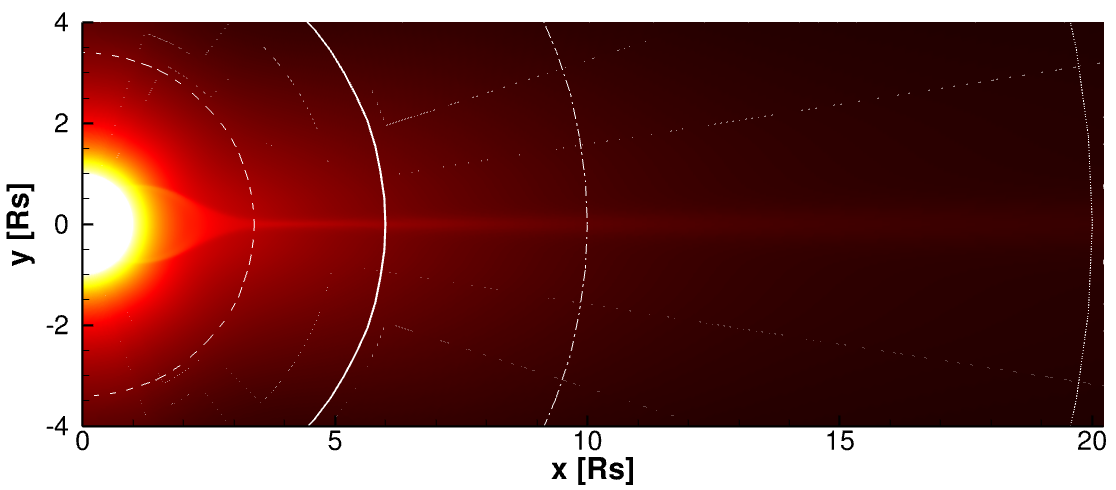}
\caption{Snapshot of logarithmic density of the variant
B2T1.32. The white lines represent slits at various distances. }\label{fig:simulation}
\end{figure}

Three simulation variants from \citet{Sorokina24} are investigated here: B1T1.32SSW, B2T1.32 and B4T1.32FSW (in notations of this paper; see Table \ref{table:cases}). We consider the 1D profiles of density, temperature, velocity and magnetic field (neglecting the small components of $\pmb{v}$ and $\pmb{B}$ tangential to the slit) at a certain height after the simulations reach a steady state. For each simulation, the height is taken at 6 $R_\odot$. To account for sub- and super-Alfv\'enic background flow speed (Table \ref{table:cases}, last column) and more accurately compare to the numerical study for the variant B2T1.32, we also consider two additional heights of 3.4 and 10 $R_\odot$ (see Table \ref{table:cases} and Fig.~\ref{fig:simulation}). The wavenumber $k$ is determined from the streamer wave wavelength, where we assume the base wavelength of 6 $R_\odot$ found in \citet{Decraemer20}. We vary this value for different slit heights to account for the wavelength growth found in \citet{Chen10,Sorokina24}. The parameters of all cases are summarised in Table \ref{table:cases}.

\begin{table}
\caption{Simulation variants.}
\label{table:cases}
\centering  
\vspace*{2mm}
\begin{tabular}{l c c c c}
\hline\hline 
Variant & $r_\mathrm{slit}$ [$R_\odot$] & $\lambda$ [$R_\odot$] & $k$ $\left[\frac{\mathrm{rad}}{R_\odot}\right]$ & $v_0(s=0)$ \\ 
\hline  
   B1T1.32SSW & 6.0 & 6.0 & 1.05 & $< v_{Ae}$\\ 
   B2T1.32$_{3.4}$& 3.4 & 5.0 & 1.25 & $< v_{Ae}$ \\
   B2T1.32& 6.0 & 6.0 & 1.05 & $> v_{Ae}$ \\
   B2T1.32$_{10}$& 10.0 & 7.0 & 0.89 & $> v_{Ae}$ \\
   B4T1.32FSW & 6.0 & 6.0 & 1.05 & $> v_{Ae}$ \\
\hline                  
\end{tabular}
\tablefoot{The key parameters are given for each simulation variant used in the \legolas{} case study. The notations are the same as in \citet{Sorokina24}. The subscripts $_{3.4}$ and $_{10}$ for the variant B2T1.32 correspond to additional slit heights.}
\end{table}

For simulation B2T1.32, the profiles along the three slits at 3.4, 6, and 10 $R_\odot$ are shown in Fig~\ref{fig:profiles}. Strictly speaking, these states are not exactly force-balanced, but the effect of this discrepancy on the frequencies of interest is assumed to be limited, such that it does not significantly affect the results of this study. A comparison of an unbalanced and balanced state is presented in App. \ref{app}.

\begin{figure}[hpt!]
\includegraphics[width=0.4\textwidth]{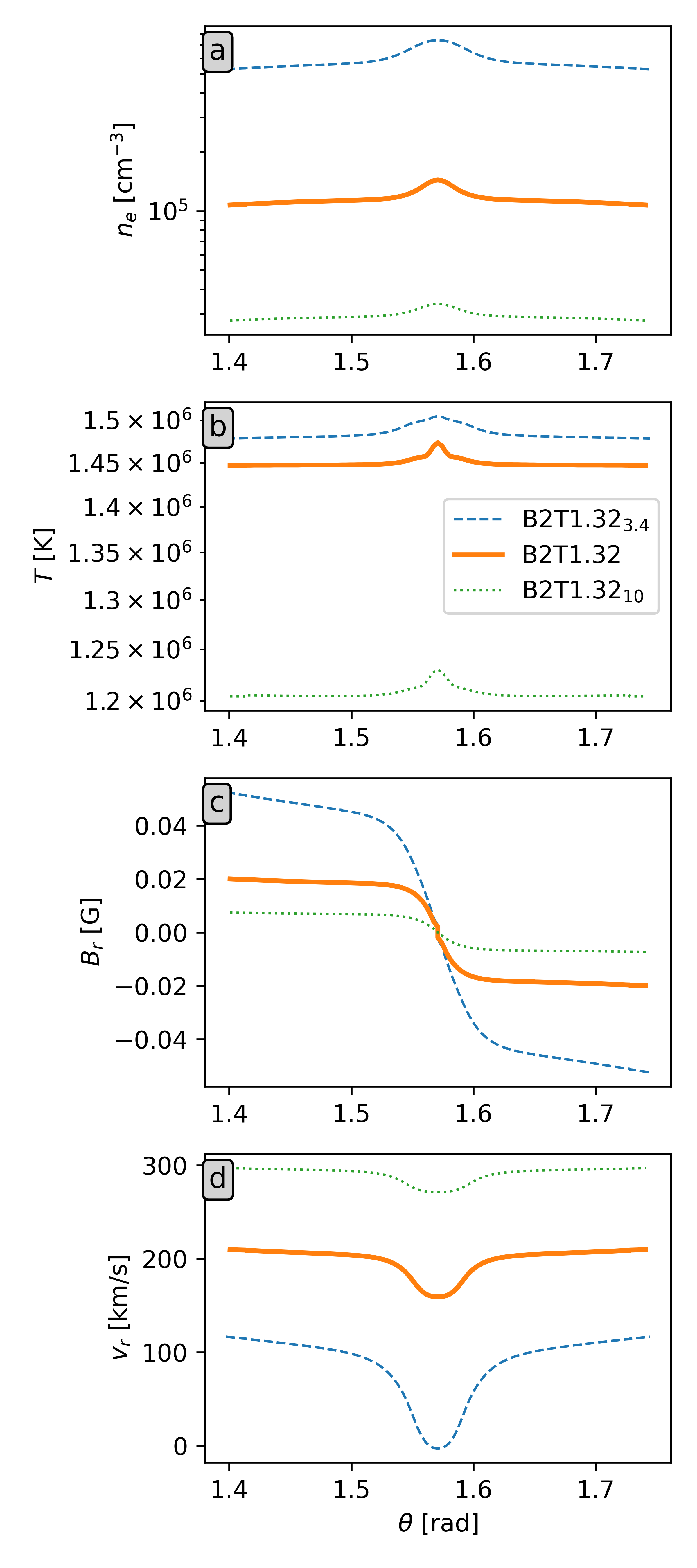}
\caption{The profiles of the quantities (a) $n$, (b) $T$, (c) $B_r$, and (d) $v_r$ along the slits in Fig.~\ref{fig:simulation} for simulation B2T1.32. Bold orange lines correspond to the slit height of 6 $R_\odot$.}
\label{fig:profiles}
\end{figure}

To ensure that the wall boundary conditions are sufficiently far away, all profiles were extrapolated as functions of $s$ using quadratic or cubic splines, to the nearest minimum/maximum on either side. The resulting profiles are shown in Fig. \ref{fig:modprofiles}. Then, for each case, wall boundary conditions were imposed at $\pm 10\max\{|s^{(e)}|\}$ where $s^{(e)}$ are the locations of the modified profiles' edges. Between each profile's edge and the wall, a quantity's value was set to the associated extrapolated profile's nearest edge value.

\begin{figure}[hpt!]
\includegraphics[width=0.4\textwidth]{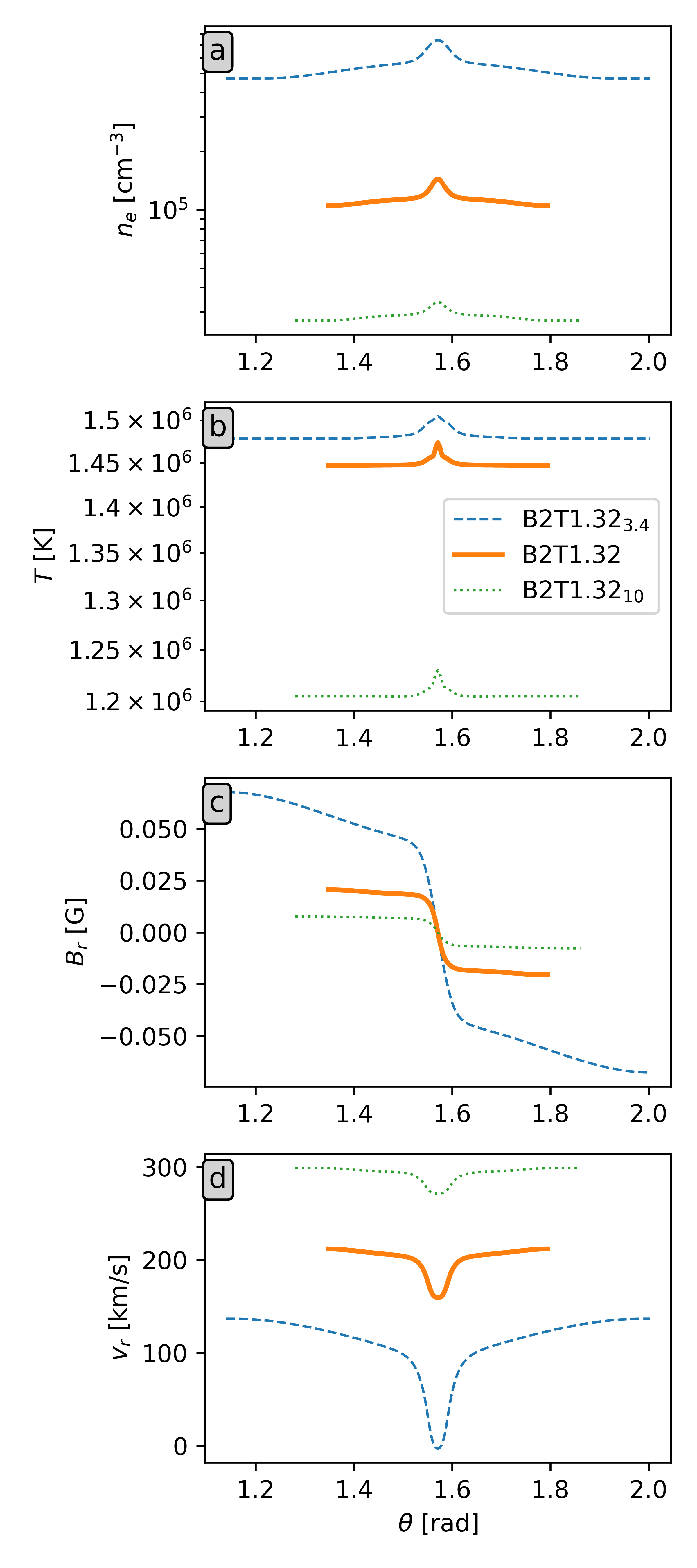}
\caption{The modified profiles of the quantities (a) $n$, (b) $T$, (c) $B_r$, and (d) $v_r$ along the slits in Fig.~\ref{fig:simulation} for the case B2T1.32. The subscripts ${3.4}$ and ${10}$ correspond to the slit heights. Bold lines correspond to the slit height of 6 $R_\odot$.}
\label{fig:modprofiles}
\end{figure}

Finally, the resulting profiles were normalised with a unit length of $L_0 = 1\,R_\odot$, and unit magnetic field $B_0$ and temperature $T_0$ equal to their respective profiles maximum values. All other unit quantities were derived assuming a fully ionised hydrogen plasma.

\section{Results: spectroscopic analysis}

In our analysis with the \legolas{} code, we aim to identify streamer waves as eigenmodes of a streamer slab, more specifically as fast body kink modes \citep{Chen10, Feng11, Decraemer20}. This means that the streamer wave is a fast magnetoacoustic wave with a displacement transverse to the streamer slab, implying that we are looking for a mode in the fast (i.e. outermost) sequence of the spectrum with an antisymmetric density eigenfunction. Before turning to the analysis of the simulation data, we illustrate this using the Epstein configuration.

\subsection{Epstein profile}\label{sec:legolas-epstein}
Supplementing the Epstein profile, Eq. \ref{eq:epstein}, and the uniform magnetic field $\pmb{B} = B_0\,\pmb{e}_z$ with a temperature $T(x) = p_0 / \rho(x)$, we obtain a static, force-balanced state, which we implement in \legolas{}. Setting the (dimensionless) parameters to $\rho_\infty = 0.5$, $\rho_0 = 1$, $p_0 = 1$, $B_0 = 10$, and $a = 0.25$, the plasma-$\beta$ is $\beta = 0.02 \ll 1$, satisfying the assumption of the analytical solution obtained by \citet{Nakariakov95}. Introducing walls at $x_w = \pm 10$ we perform $100$ \legolas{} runs on a uniform grid with $201$ grid points with $ka$ ranging from $0.025$ to $2.5$. Since all eigenmodes have purely real frequencies, the results can be combined into a dispersion diagram. This is done for the first $25$ runs in Fig. \ref{fig:epstein}a, where we included the analytical solution, Eq. \ref{eq:epstein-dispersion}, and marked the kink waves with crosses. The kink waves were identified manually by their antisymmetric density (Fig. \ref{fig:epstein}c) and peaked $v_x$ (Fig. \ref{fig:epstein}d) perturbation amplitudes. Note that the beginning of the fast wave sequence, and thus the fast body kink wave, overlaps with the Alfv\'en continuum, which arises due to the continuous variation in Alfv\'en speed throughout the non-uniform slab. To illustrate this, the Alfv\'en continuum has been shaded grey in the figure.

\begin{figure}
    \centering
    \includegraphics[width=\linewidth]{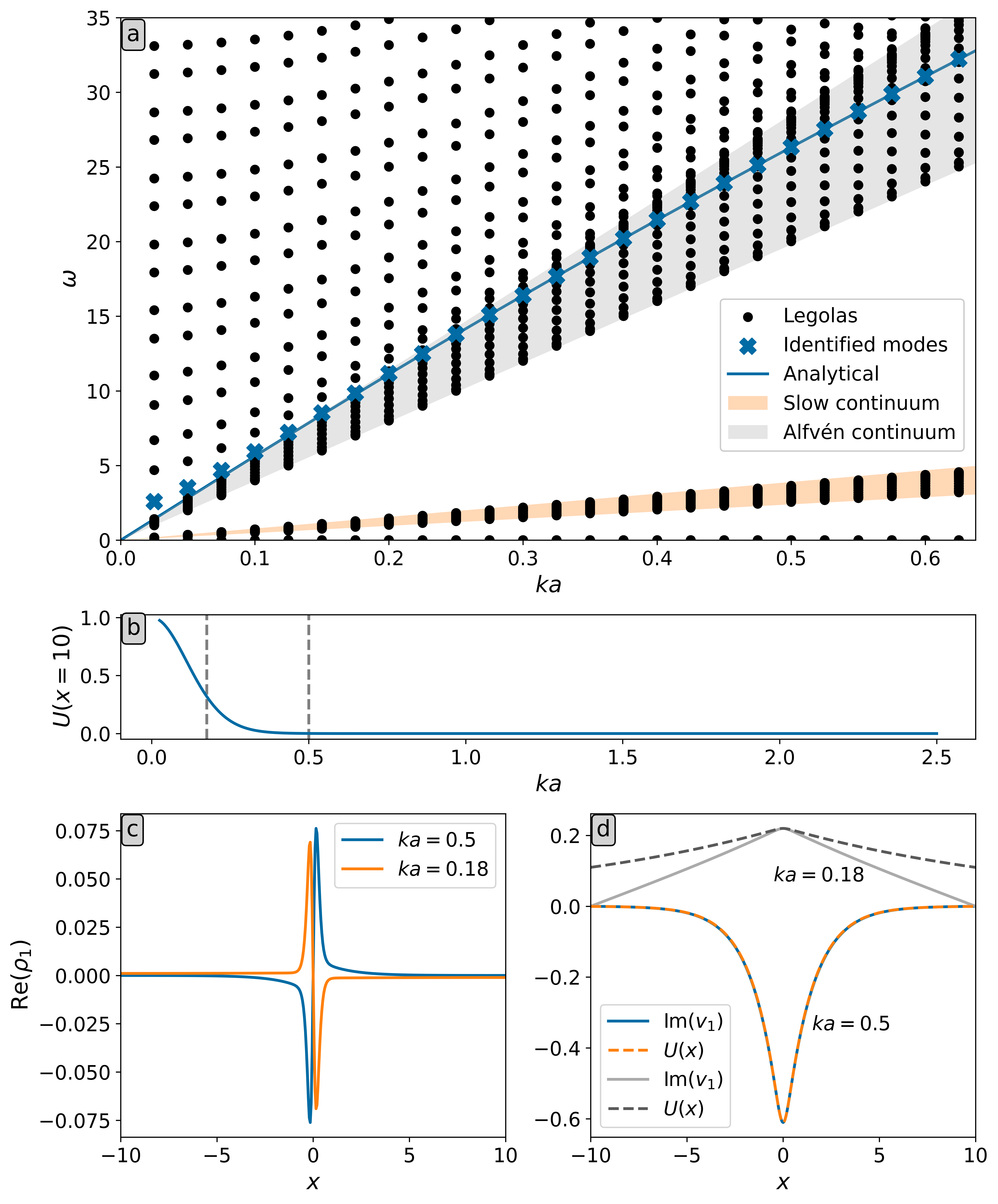}
    \caption{(a) Dispersion diagram for the Epstein profile combining the first $25$ \legolas{} runs. The analytical kink wave frequency is shown as a solid line and the identified \legolas{} modes are marked with crosses. The slow and Alfv\'en continua are shaded. (b) Eq. \ref{eq:epstein-U} evaluated at the wall position $x_w = 10$ for varying wavenumber $ka$. (c) Density perturbation amplitude of the \legolas{} kink mode for $ka=0.18$ and $ka=0.5$. (d) $v_x$ perturbation amplitude of the \legolas{} kink mode for $ka=0.18$ and $ka=0.5$ compared to the analytical expression $U(x)$.}
    \label{fig:epstein}
\end{figure}

\legolas{} and the analytical solution agree well for the entire range, except for a deviation observed for small $ka$. This deviation is due to the incompatibility of \legolas{}'s boundary conditions with the analytical solution. Note that the profile $U(x)$ widens as the exponent $\nu$ in Eq. \ref{eq:epstein-U}, and thus $ka$, decreases. Evaluating $U(x)$ at the wall position $x_w = 10$ for various $ka$ reveals that it deviates significantly from zero, as shown in Fig. \ref{fig:epstein}b, such that the kink mode is strongly affected by the $v_1=0$ boundary condition in \legolas{}. Indeed, extending the interval in \legolas{}, i.e. moving the walls further out, shifts the kink frequency towards the analytical value. Despite this deviation, these modes can still be considered kink waves as attested by the density perturbations shown in Fig. \ref{fig:epstein}c, for $ka=0.5$ (analytical agreement) and $ka=0.18$ (deviation from analytical result), whose $U(x_w)$ values have been indicated with dashed lines in Fig. \ref{fig:epstein}b. Finally, Fig. \ref{fig:epstein}d shows how the $v_x$ perturbation amplitude in \legolas{} coincides with $U(x)$ for $ka=0.5$, but is modified from the analytical profile for $ka=0.18$.

Hence, we conclude that \legolas{} is well-suited to calculate the streamer slab's fast body kink waves for sufficiently large wavenumbers, or alternatively, for sufficiently extended intervals around the inhomogeneity.

\subsection{Simulation data}
In Figs. \ref{fig:B2T1-instability}-\ref{fig:B2T1-1pair}a we show the resulting spectrum for case B2T1.32 for $501$ grid points. Each marker corresponds to an eigenfrequency $\omega$ in the complex plane, including the (discreetly sampled) Alfv\'en and slow continua, and the propagating up- and downstream discrete modes of the fast wave sequences. As indicated in Fig. \ref{fig:B2T1-instability}a, the asymmetry of the spectrum with respect to $\mathrm{Re}(\omega)=0$ is caused by a Doppler shift to the right due to the background flow.

\begin{figure}
\includegraphics[width=0.5\textwidth]{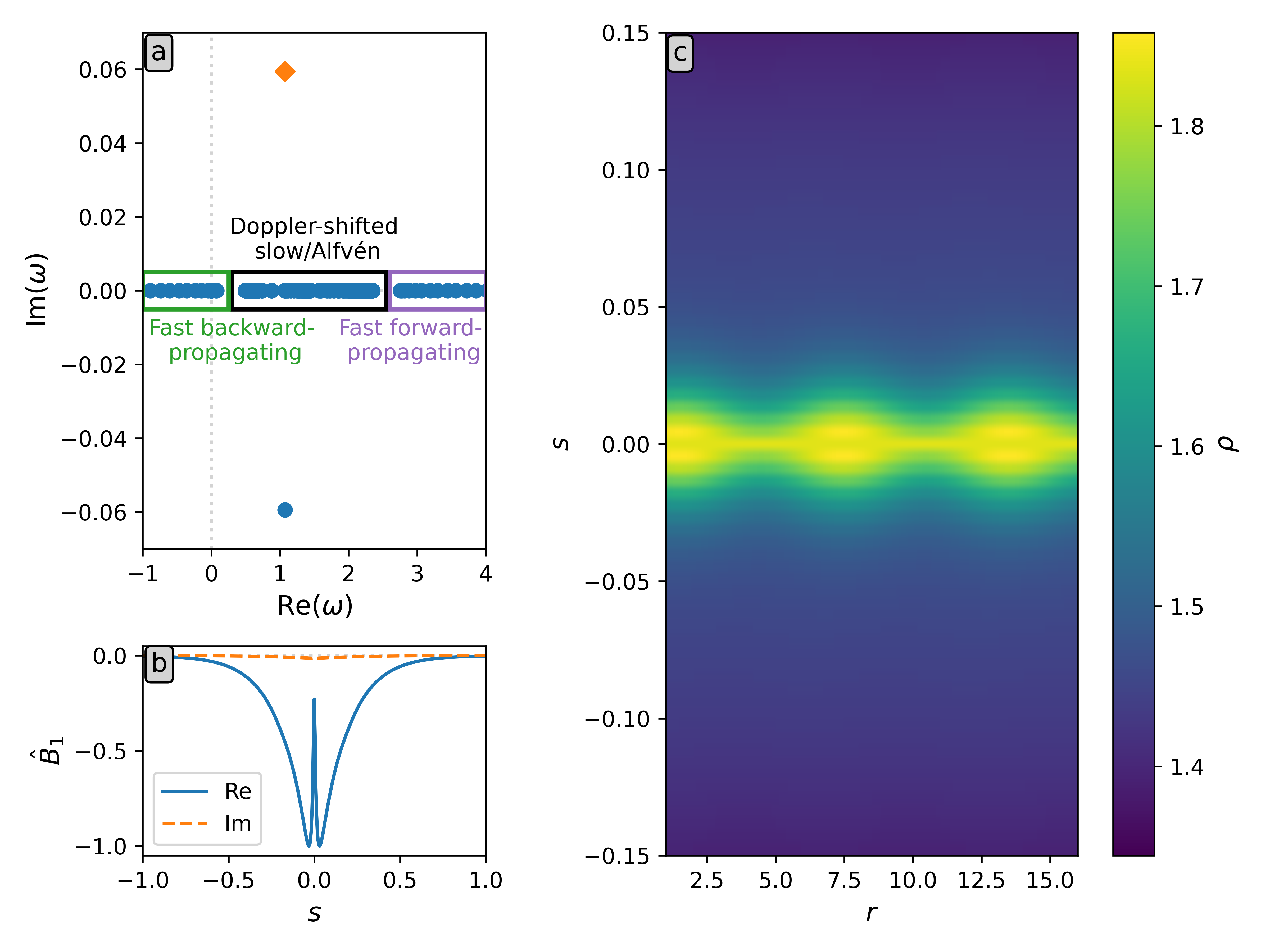}
\caption{(a) Spectrum of the variant B2T1.32 at $r = 6\ R_\odot$ for $\lambda = 6\ R_\odot$. The orange diamond marks the unstable tearing-like mode. Eigenmode continua and sequences have been boxed and labelled. (b) The $\hat{B}_1$ perturbation amplitude of the tearing-like mode. (c) The resulting slab perturbation.}
\label{fig:B2T1-instability}
\end{figure}

\begin{figure}
\includegraphics[width=0.5\textwidth]{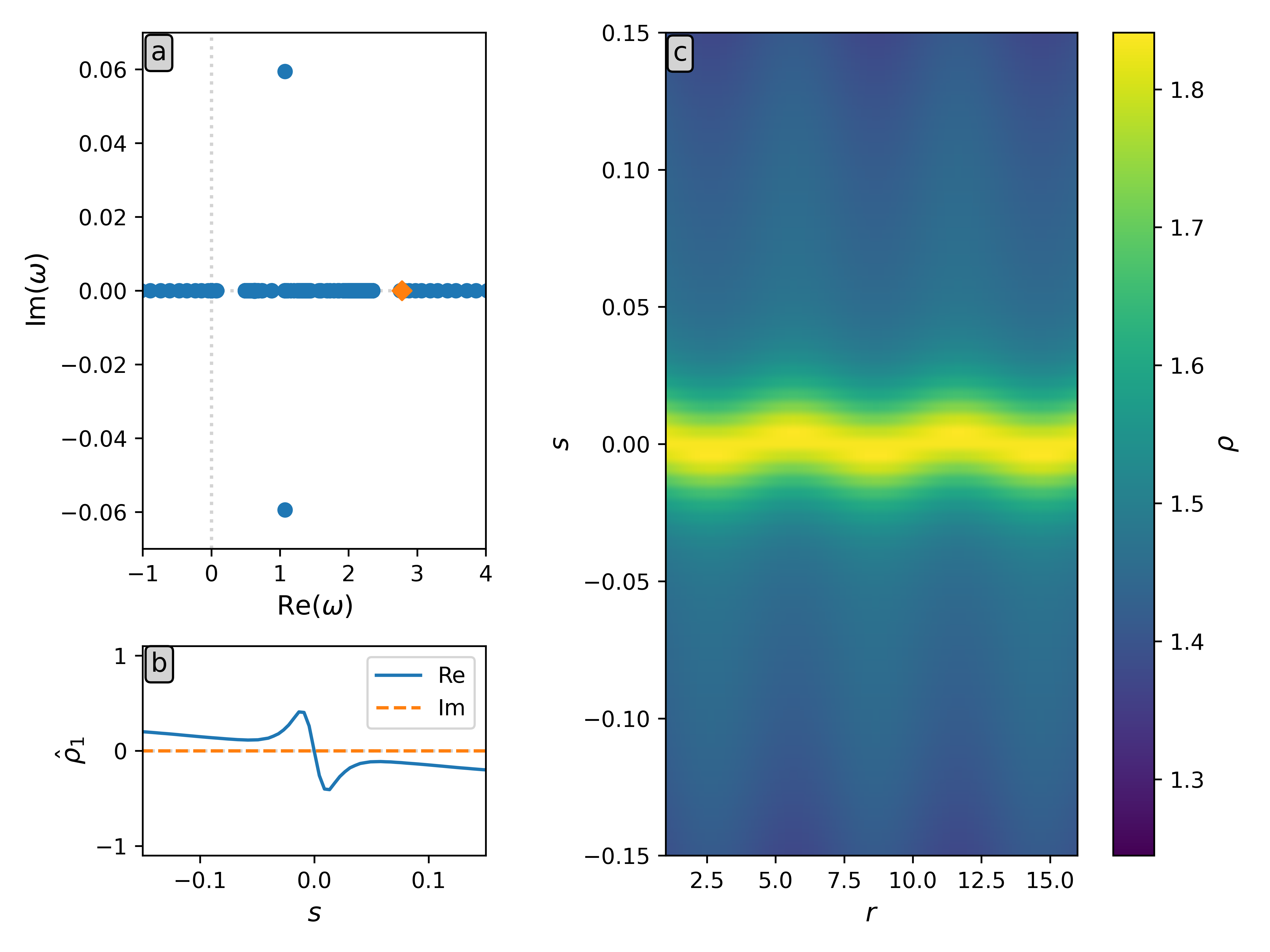}
\caption{(a) Spectrum of the variant B2T1.32 at $r = 6\ R_\odot$ for $\lambda = 6\ R_\odot$. The orange diamond corresponds to the first forward-propagating discrete mode in the sequence. (b) The $\hat{\rho}_1$ perturbation amplitude of the marked mode. (c) The resulting slab perturbation.}
\label{fig:B2T1-1mode}
\end{figure}

\begin{figure}
\includegraphics[width=0.5\textwidth]{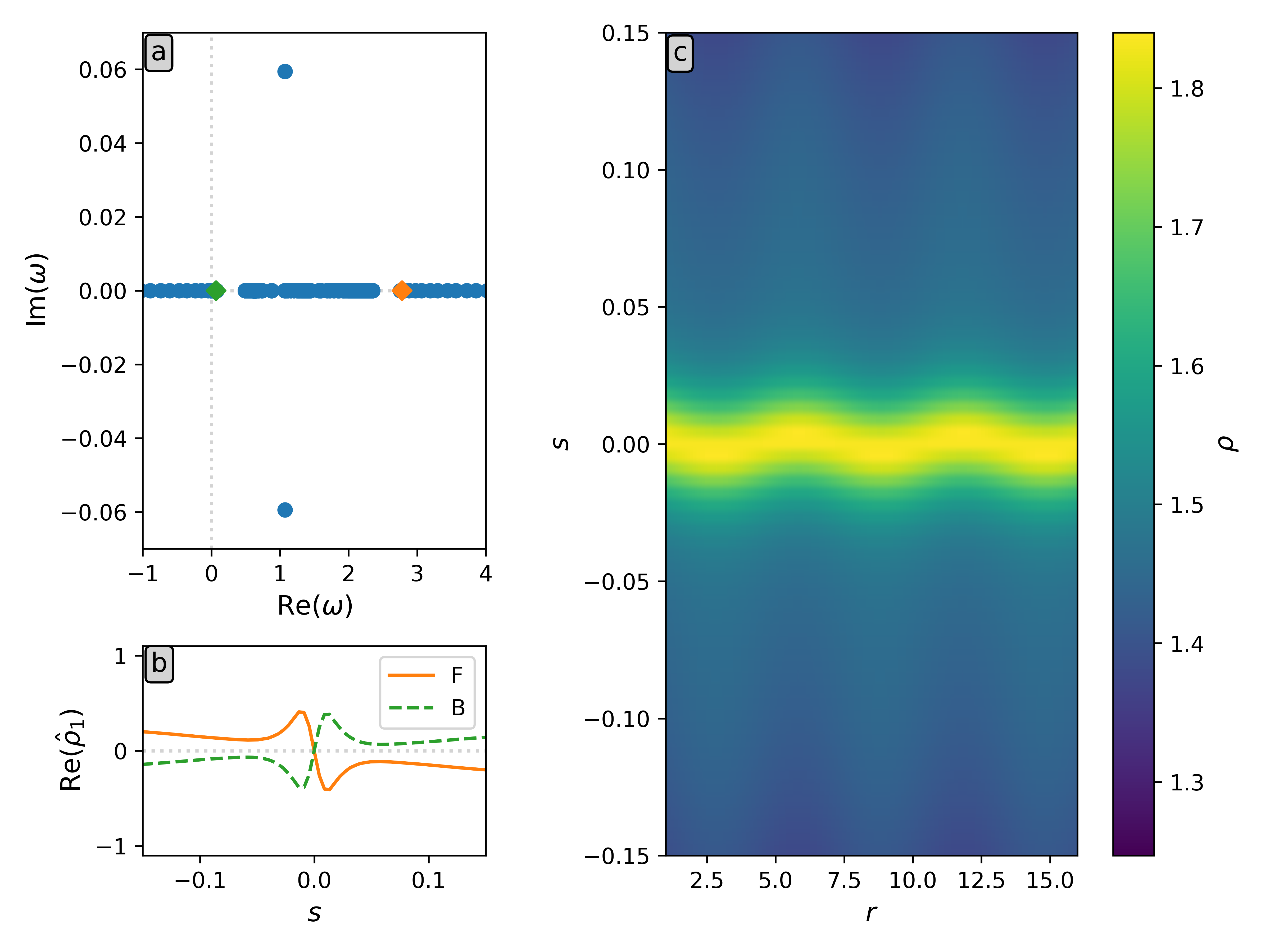}
\caption{(a) Spectrum of the variant B2T1.32 at $r = 6\ R_\odot$ for $\lambda = 6\ R_\odot$. The orange and green diamonds mark the first Doppler-shifted forward (F) and backward (B) propagating fast body mode, respectively. (b) The $\hat{\rho}_1$ perturbation amplitudes of the marked modes. (c) The resulting slab perturbation.}
\label{fig:B2T1-1pair}
\end{figure}

Though not the focus of this work, we first comment on the unstable mode, shown as an orange diamond in Fig.~\ref{fig:B2T1-instability}a. Since its $\hat{B}_1$-component (Fig. \ref{fig:B2T1-instability}b) is non-zero whilst its other magnetic perturbation components are simultaneously zero at $s=0$, this indicates magnetic reconnection across the sheet, and thus this is a tearing-like instability propagating with the background flow speed of 160 km/s inside the slab (see Fig.~\ref{fig:profiles}d). The typical island structure of the tearing mode is recovered in the 2D spatial view of Fig. \ref{fig:B2T1-instability}c. Considering resistivity was not included, we would not expect the presence of a tearing instability. However, the tearing instability could be part of the operator's pseudospectrum, only becoming a true solution with a small resistivity, such that it appears here due to numerical inaccuracies, presumably introduced by the numerical derivatives of the simulation profiles.

Now, let us consider the first forward-propagating discrete mode in the sequence (orange diamond in Fig. \ref{fig:B2T1-1mode}a).\footnote{Actually, the mode shown is the second mode in the sequence calculated by \legolas{}. The first solution has a strictly positive density perturbation amplitude everywhere, i.e. it has an `effective vertical wavenumber' of zero \citep[no nodes; see e.g.][]{Goedbloed19}, and is thus called the zeroth mode.} Fig.~\ref{fig:B2T1-1mode}b shows the corresponding density perturbation amplitude, typical of a kink mode \citep{Harris62}. Finally, Fig. \ref{fig:B2T1-1mode}c shows the matching 2D spatial view of the slab oscillation. We may thus assume that the streamer wave may be recreated with the first forward-propagating mode in the fast wave sequence. This supports the idea that the streamer wave is an eigenmode of the streamer slab and may be approximated as a fast body kink mode, confirming the potential of using such waves for seismology.

We also consider the inclusion of a Doppler shifted backward-propagating kink mode (note: the Doppler shift of the backward propagating mode results in $\omega, v_{ph} > 0$), i.e\ we consider the pair consisting of the first mode in each fast sequence (orange and green diamonds in Fig.~\ref{fig:B2T1-1pair}a). Their density perturbation amplitudes are shown in Fig. \ref{fig:B2T1-1pair}b and the 2D spatial view in Fig. \ref{fig:B2T1-1pair}c. To distinguish between this superposition and the forward propagating mode only, Fig.~\ref{fig:timeplot} shows the comparison of their temporal evolutions. On the left, we show the perturbed slab with an indication of where the diagnostic cut is taken. On the right, we show the resulting space-time diagrams. In case of the first forward-propagating mode (top row), the space–time diagram shows diagonal streaks: each crest shifts radially outward in time, corresponding to a kink-like travelling front. In contrast, the pair of forward-backward propagating modes (bottom row) shows maxima and minima flip periodically. We observe that for the latter the waves still propagate forward along the slab, but this results in an oscillatory pattern with nodes (points of zero displacement) and anti-nodes (points of maximum displacement). Thus the pair produces standing wave behaviour rather than a forward-propagating kink wave.

   \begin{figure}[hpt!]
   \includegraphics[width=0.5\textwidth]{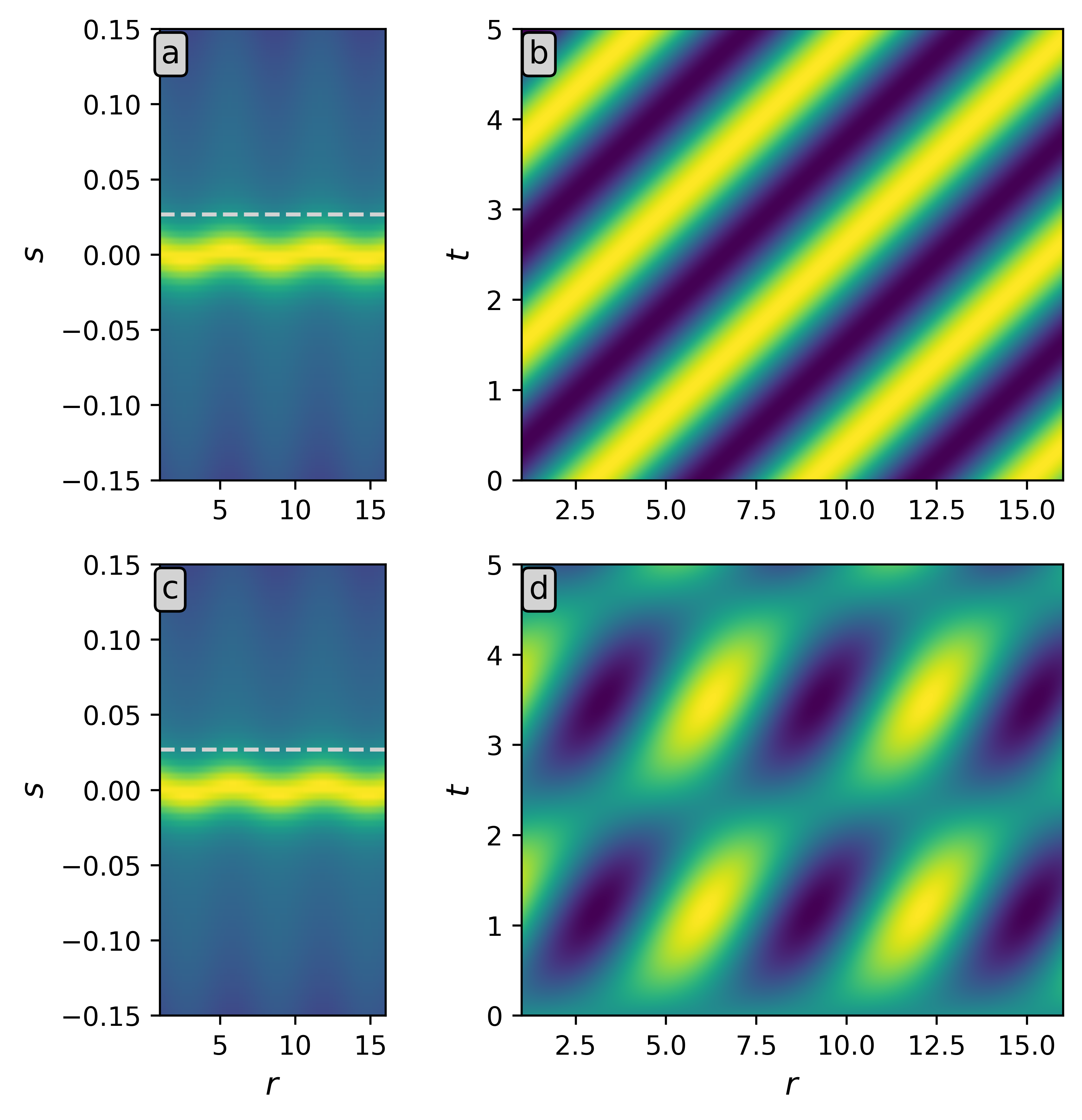}
   \caption{Comparison between the evolution of a slab for the two cases, considered in Fig.~\ref{fig:B2T1-1mode} and Fig.~\ref{fig:B2T1-1pair}, for variant B2T1.32. (a) Slab perturbation of the first forward-propagating mode. (c) Slab perturbation of the first pair of forward-backward propagating modes. The horizontal dashed lines mark the $s$-slice used to construct the space–time diagrams in (b) for the forward propagating mode and (d) for the pair of forward-backward propagating modes.}
         \label{fig:timeplot}
   \end{figure}

Finally we note that unlike the variant B2T1.32 shown so far, not all cases have clearly separated Alfv\'en and fast sequences. For the five variants listed in Table \ref{table:cases}, the spectrum in the vicinity of the beginning of the forward-propagating fast sequence is shown in Fig. \ref{fig:identified}, with the identified fast body kink mode marked with an orange diamond in each spectrum. For variants B2T1.32, B2T1.32$_{10}$, and B4T1.32FSW, the Alfv\'en and fast sequences are visibly distinct, but for B1T1.32SSW and B2T1.32$_{3.4}$ they overlap. In these cases, the mode of interest was identified by comparing their density and $v_x$ perturbation profiles to the ones found for the Epstein profile (see Sec. \ref{sec:legolas-epstein}).

\begin{figure}
\centering
\includegraphics[width=0.5\textwidth]{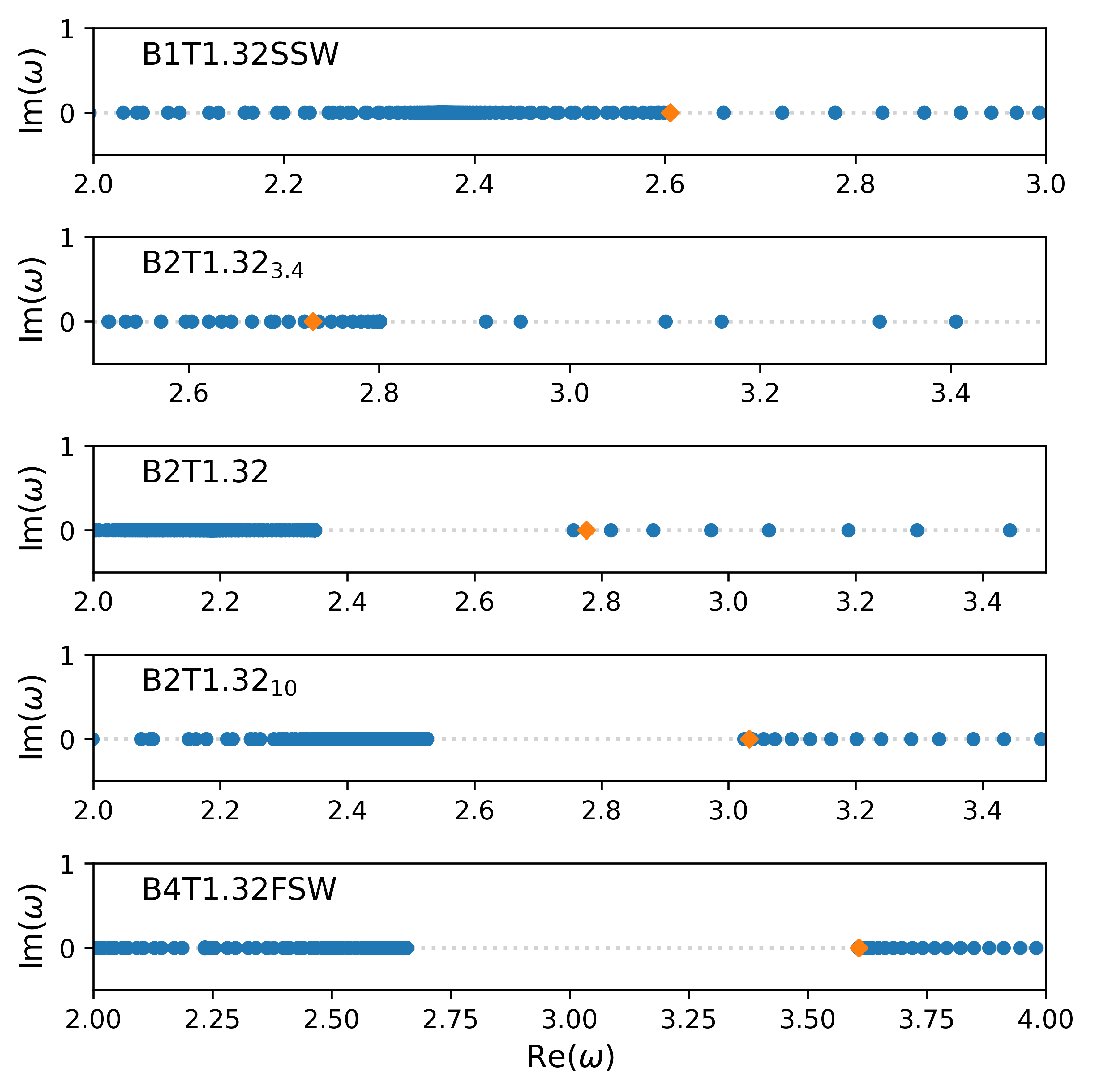}
\caption{Zoom of each case's spectrum with the identified forward-propagating kink mode marked with an orange diamond.}
      \label{fig:identified}
\end{figure}

The analysis of the eigenfunctions of a streamer slab may be extended further, however such considerations are beyond the scope of this article. Going forward, we focus solely on the properties of the identified forward-propagating kink waves.

\section{Results: comparison to analytical theory, simulations, and observations}

To study whether the selected modes are compatible with streamer wave observations, we first consider their phase speeds. Table \ref{table:vph} lists the frequencies and phase speeds $v_{ph} = \omega/k$ obtained with \legolas{} for each variant, as well as the equivalent mode when neglecting background flow (subscript $0$). In the absence of flow, the spectrum is symmetric about the imaginary axis (no Doppler shift).

\begin{table}
\caption{\legolas{} kink wave phase speeds.}
\label{table:vph}
\centering  
\vspace*{2mm}
\small
\begin{tabular}{l c c c c }
\hline\hline 
Variant & $\omega$ $\left[\frac{\mathrm{rad}}{\mathrm{s}}\right]$ & $\omega_0$ $\left[\frac{\mathrm{rad}}{\mathrm{s}}\right]$ & $v_{ph}$ $\left[\frac{\mathrm{km}}{\mathrm{s}}\right]$ & $v_{ph,0}$ $\left[\frac{\mathrm{km}}{\mathrm{s}}\right]$ \\ 
\hline  
   B1T1.32SSW & $5.0\times 10^{-4}$ & $2.8\times 10^{-4}$ & $330$ & $188$ \\ 
   B2T1.32$_{3.4}$ & $6.2\times 10^{-4}$ & $3.8\times 10^{-4}$ & $342$ & $210$ \\ 
   B2T1.32 & $6.2\times 10^{-4}$ & $3.0\times 10^{-4}$ & $413$ & $202$ \\   
   B2T1.32$_{10}$ & $6.2\times 10^{-4}$ & $2.4\times 10^{-4}$ & $481$ & $183$ \\
   B4T1.32FSW & $1.2\times 10^{-3}$ & $4.4\times 10^{-4}$ & $786$ & $295$ \\  
\hline                  
\end{tabular}
\tablefoot{The kink wave frequencies and their phase speeds $v_{ph} = \omega/k$ as obtained with \legolas{} including and excluding (subscript $0$) background flow.}
\end{table}

Let us consider the central reference case B2T1.32. For this case the Mach number $M_{i/e}=\frac{v_{i/e}}{c_{si/e}}$ is 1.28/1.55 and the Alfv\'enic Mach number $M_{Ai/e}=\frac{v_{i/e}}{v_{Ai/e}}$ is 2.83/1.28. Thus, we assume both regions are mildly to moderately supersonic ($M>1$) and super-Alfv\'enic ($M_A>1$) and kink waves are strongly advected downstream, more compressible inside, and relatively stabilised by external magnetic tension against moderate shear. In Fig.~\ref{fig:speeds} we compare the characteristic speeds $v_{A}$, $v_{i/e}=v_r=v_{SW}$ (where $SW$ refers to the background solar wind), $c_{T}$ found in the \amrvac{} simulation B2T1.32, and the phase speeds $v_{ph} = \omega/k$ obtained for that case with the various methods. The \legolas{} phase speed is slightly higher than the one estimated from the \amrvac{} simulation in \citet{Sorokina24} and than we would expect from the analytical approximation $v_{Ae} + v_e$, Eq. \ref{eq:anal-expected}.

A deviation of the \legolas{} result from the \amrvac{} estimate is not surprising, considering several approximations were made in the \legolas{} analysis. The \amrvac{} measurement takes the inhomogeneity perpendicular to the slit into account, as well as solar gravity in that direction, both of which were neglected in \legolas{}. The spectroscopic analysis also discarded the small $B_{\theta}$- and $v_{\theta}$-components in \amrvac{}. Finally, the external velocity in the \legolas{} study was slightly higher than in \amrvac{} due to the extrapolation in pre-processing (see Sec. \ref{sec:setup}). Despite these differences, \legolas{} obtains a phase speed of comparable magnitude to \amrvac{}.

   \begin{figure}
   \centering
   \includegraphics[width=0.4\textwidth]{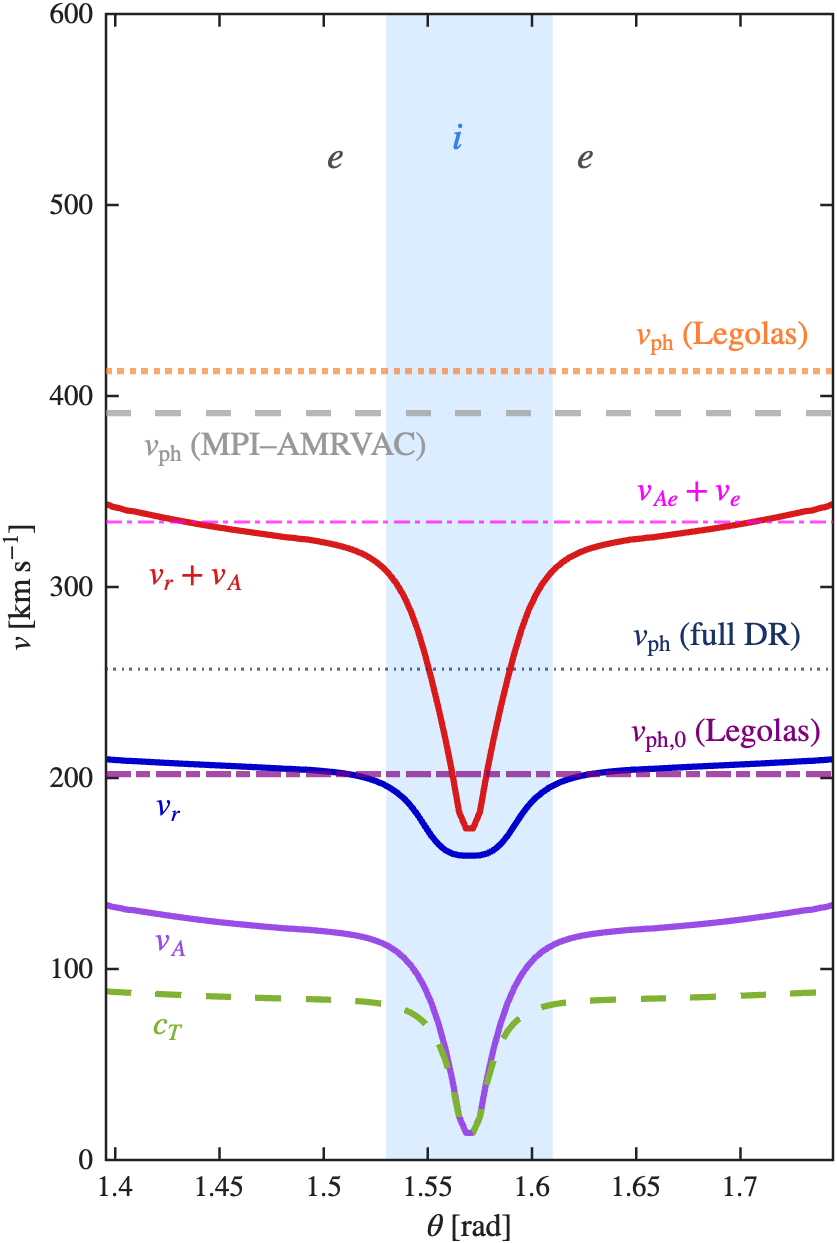}
   \caption{The characteristic speeds inside and outside the streamer slab for variant B2T1.32 ($r = 6$ $R_\odot$). The horizontal lines represent the phase speeds obtained using various methods.}
         \label{fig:speeds}
   \end{figure}

For the analytical approximation, note that it is derived under the assumption that $ka \ll 1$, whereas the estimate of $ka$ for variant B2T1.32 is around $0.25$ for a slab thickness as indicated in Fig. \ref{fig:speeds}, with even larger values for other variants. However, the same result also holds if we abandon the slender slab approximation and consider a full transcendental dispersion relation in presence of a background flow, Eq. \ref{eq:full-DR}. Evaluating this dispersion relation for the parameters of variant B2T1.32, using $a = 0.2\ R_{\odot}$ and internal values at position $s_i = 0.118\ R_{\odot}$ and external values at $s_e = 0.805\ R_{\odot}$, we find a phase speed of $257$ km/s (indicated in Fig. \ref{fig:speeds} as `$v_{\mathrm{ph}}$ (full DR)'). For this slab thickness the full dispersion relation already deviates from the simplified formula $v_{ph}=v_e+v_{Ae}$, which yields $334$ km/s for this choice of $s_i, s_e$, and even further from the \legolas{} result (see Fig.~\ref{fig:speeds}). Considering the profiles here are smooth rather than discontinuous as assumed in the analytical model, we assess the influence of our choice of $s_e, s_i$ by varying one whilst keeping the other fixed. This is shown in Fig. \ref{fig:vph-analytical}. Though the role of $s_e$ is extremely limited, the result depends strongly on $s_i$. This makes a clear comparison of analytical to numerical results impossible, and we are thus unable to validate the relation $v_{ph}=v_e+v_{Ae}$ for continuously varying background profiles.

\begin{figure}
    \centering
    \includegraphics[width=\linewidth]{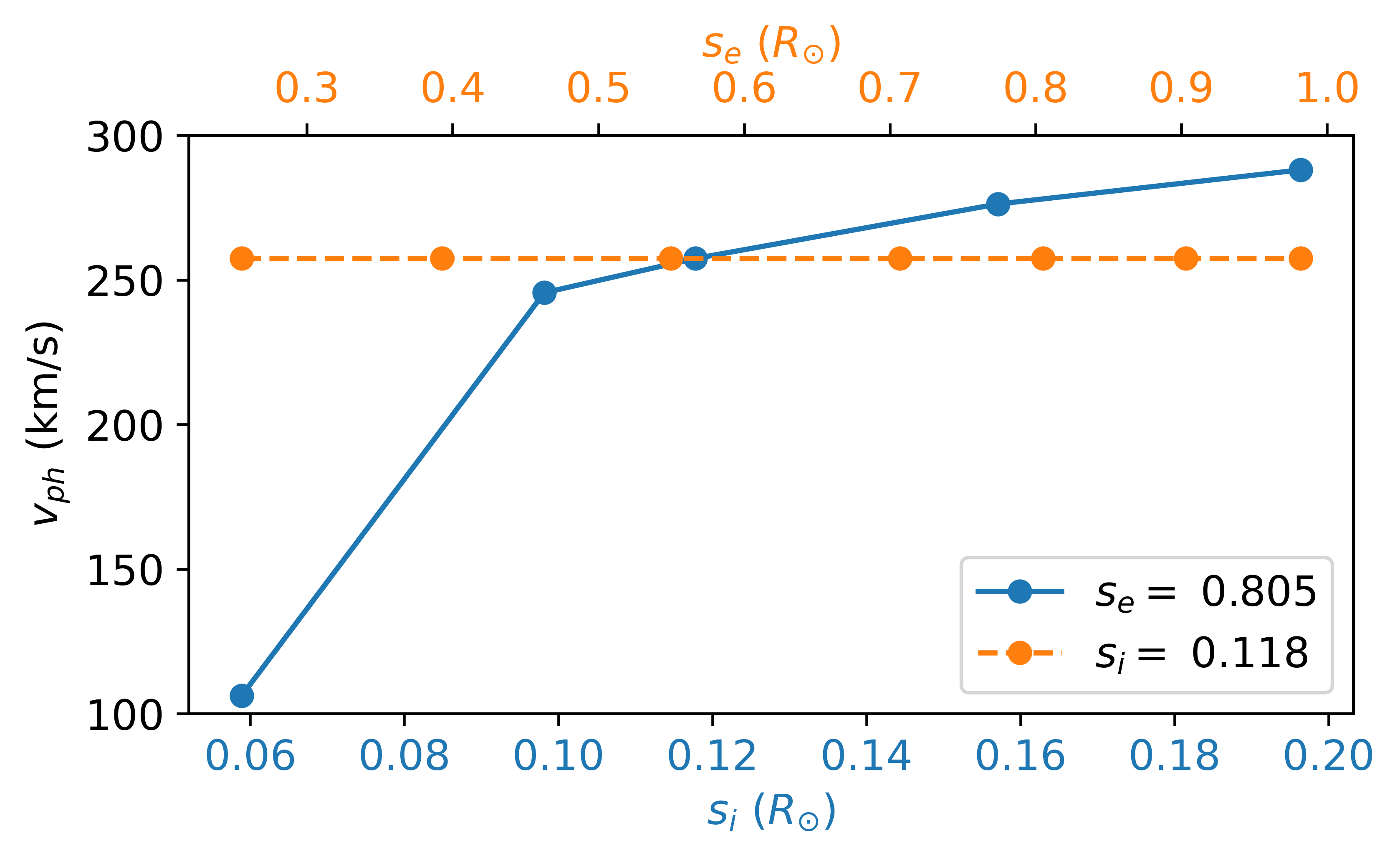}
    \caption{Solution of Eq. \ref{eq:full-DR} for $v_{ph}$ for different combinations of the internal ($s_i$) and external ($s_e$) locations where the values were extracted from the profiles of variant B2T1.32.}
    \label{fig:vph-analytical}
\end{figure}

To compare to the numerical \amrvac{} study \citep{Sorokina24} in more detail, we consider the lab-frame phase speed $v_{ph}$ and the rest-frame phase speed $v_{ph,0}$ (omitting background flow; see Table \ref{table:vph}). In Fig.~\ref{fig:vph0}, $v_{ph,0}$ vs. $v_{ph}$ results are shown for both \amrvac{} and \legolas{}, and the speed expected from the analytical expression if we operate with the averaged values of external Alfv\'en speed. Overall, as already mentioned for both numerical studies we observe that the measured phase speed is noticeably different from the analytical estimation $v_{ph}=v_{SW}+v_{Ae}$, except for the slow solar wind (SSW) variant.

   \begin{figure}
   \centering
   \includegraphics[width=0.4\textwidth]{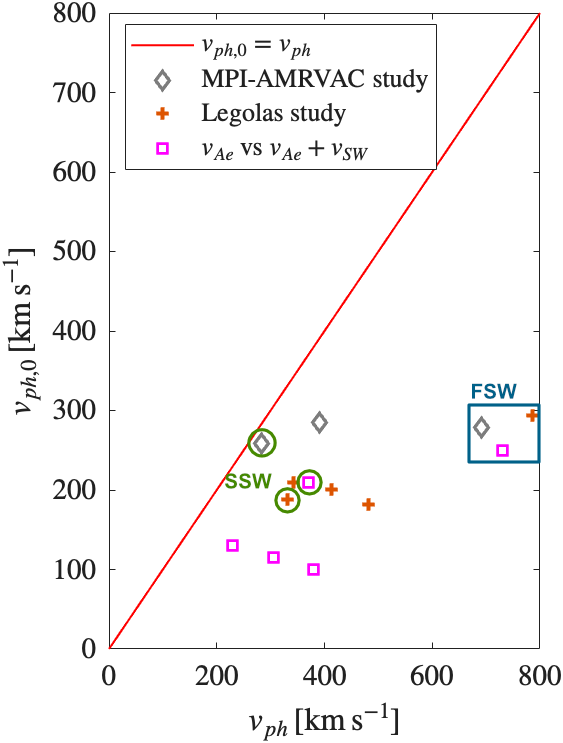}
   \caption{Scatter plot of measured phase speed ($v_{ph}$) vs. phase speed of the wave in plasma rest frame ($v_{ph,0}$) in the numerical studies. The results for variants B1T1.32SSW and B4T1.32FSW have been marked. The other points belong to simulation B2T1.32 at various slit heights, with slit height increasing from left to right (see also Table \ref{table:vph}).}
         \label{fig:vph0}
   \end{figure}

   \begin{figure}
   \centering
   \includegraphics[width=0.4\textwidth]{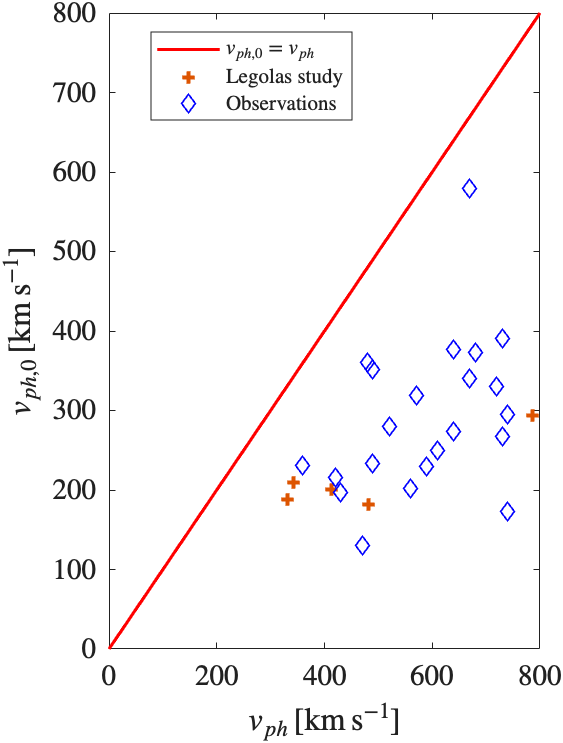}
   \caption{Scatter plot of measured phase speed ($v_{ph}$) vs. phase speed of the wave in plasma rest frame ($v_{ph,0}$) comparing the \legolas{} study with observational results.}
         \label{fig:vph0-obs}
   \end{figure}

In Fig.~\ref{fig:vph0-obs} we show the same \legolas{} results listed in Table \ref{table:vph} alongside the observational results from \citet{Decraemer20}. The phase speeds obtained from the \legolas{} study appear to be consistent with those reported in observations, though on the lower end. The \legolas{} case to the far right corresponds to a fast solar wind background (B4T1.32FSW), which is not typically associated with helmet streamer waves.

\section{Conclusions} \label{sec:conclusions}

The \legolas{} analysis verified that the streamer wave may be considered an eigenmode of the streamer slab, namely a forward propagating fast body kink mode. Therefore it is a good candidate for coronal seismology. Overall, the phase speeds found in the \legolas{} analysis are in agreement with the observational and numerical studies. However, the question of whether we can rely on the analytical approximation in the long-wave limit to estimate the solar wind speed remains, since \legolas{}'s numerical results show a discrepancy with the simple linear fit $v_{ph}=v_{SW}+v_{Ae}$ in all cases but the slow solar wind background. In this comparison, it should be noted that the assumptions made about the wavelength and slab thickness in the derivation of the simple analytical relation may not hold for the considered numerical cases, in part because the slab thickness is not well-defined for the smoothly varying profiles obtained from the simulations. Furthermore, the solution of the full analytical dispersion relation depends strongly on which internal values are considered, severely limiting the usability of the analytical result for smoothly varying profiles. This highlights that even though the mode identification was done correctly, the analytical and 1D wave mode solutions are difficult to compare to 2D simulations and 3D observations, indicating that seismology should be performed very carefully. Nevertheless, the matching case warrants further investigation, considering helmet streamers are generally associated with slow solar wind \citep{Einaudi2001, Ofman2004, Lapenta2005, Antiochos2011}. Our results also highlight the need for a more quantitatively accurate and robust observation-based 3D model to properly capture streamer properties and dynamics and verify the discussed speed distributions.

In spite of this, future work could make a first attempt to extract coronal parameters through iterative application of the \legolas{} code. If a rough shape for the slab profiles can be inferred \citep[see e.g.][for the density structure]{Decraemer19} from a combination of meridional plane and line-of-sight observations, and the streamer wave wavelength and bulk velocity can be constrained through imaging and emission spectra, iteration on the remaining parameters may yield parameter regime estimates for which the \legolas{} result is compatible with the observed streamer wave phase speed.

Overall, coupling \legolas{} with numerical models provides a valuable framework for mode identification and understanding of their propagation, despite the simplifying assumptions and required pre-processing that may result in some discrepancy between the original numerical simulations and \legolas{}. In the future this approach can be extended from the current idealised slab conditions to a more complex scenario. Introducing additional physical effects, such as resistivity, would allow for a more realistic representation of coronal conditions, and potentially enable the estimation of the damping mechanisms. Additionally, it may shed more light on the tearing(-like) mode solution, which has not been properly investigated in this study. Finally, damping could also be achieved through resonant absorption when the analysis is not limited to waves propagating radially outward, but a non-zero azimuthal wavenumber is allowed.

\section*{Data availability}
The results in this work were obtained with \legolas{} v2.3.1, available at \url{https://legolas.science}.

\begin{acknowledgements}
The authors thank the anonymous referee for their suggestions improving the paper. JDJ thanks R. Keppens for insightful discussions regarding MHD spectroscopy. JDJ acknowledges funding by the Research Foundation - Flanders (FWO) fellowship 1225625N. TVD received financial support from the Flemish Government under the long-term structural Methusalem funding program, project SOUL: Stellar evolution in full glory, grant METH/24/012 at KU Leuven. The research that led to these results was subsidised by the Belgian Federal Science Policy Office through the contract B2/223/P1/CLOSE-UP. It is also part of the DynaSun project and has thus received funding under the Horizon Europe programme of the European Union under grant agreement (no. 101131534). Views and opinions expressed are however those of the author(s) only and do not necessarily reflect those of the European Union and therefore the European Union cannot be held responsible for them. This research has made use of NASA's Astrophysics Data System Bibliographic Services.
\end{acknowledgements}

\bibliographystyle{aa}
\bibliography{bibliography}

\begin{appendix}
\section{Force-balanced spectrum}\label{app}
As noted in Sec. \ref{sec:setup}, the simulation slices do not represent a force-balanced state satisfying
\begin{equation}\label{eq:balance}
    \frac{\partial}{\partial s}\left( \rho_0 T_0 + \frac{\pmb{B}_0^2}{2} \right) = 0.
\end{equation}
To rectify this, we could calculate a new temperature profile by substituting the data's density and magnetic field profiles in Eq. \ref{eq:balance}. The result is compared to the data's temperature profile in Fig. \ref{fig:balanced}a. For either profile, the spectrum is shown in Fig. \ref{fig:balanced}b. It is clear that the spectra are extremely close and the modes of interest, and their phase speeds, do not differ in a major way. Therefore, we assume that this deviation from force-balance does not significantly affect the conclusions of this study.

\begin{figure}[ht!]
    \centering
    \includegraphics[width=0.4\textwidth]{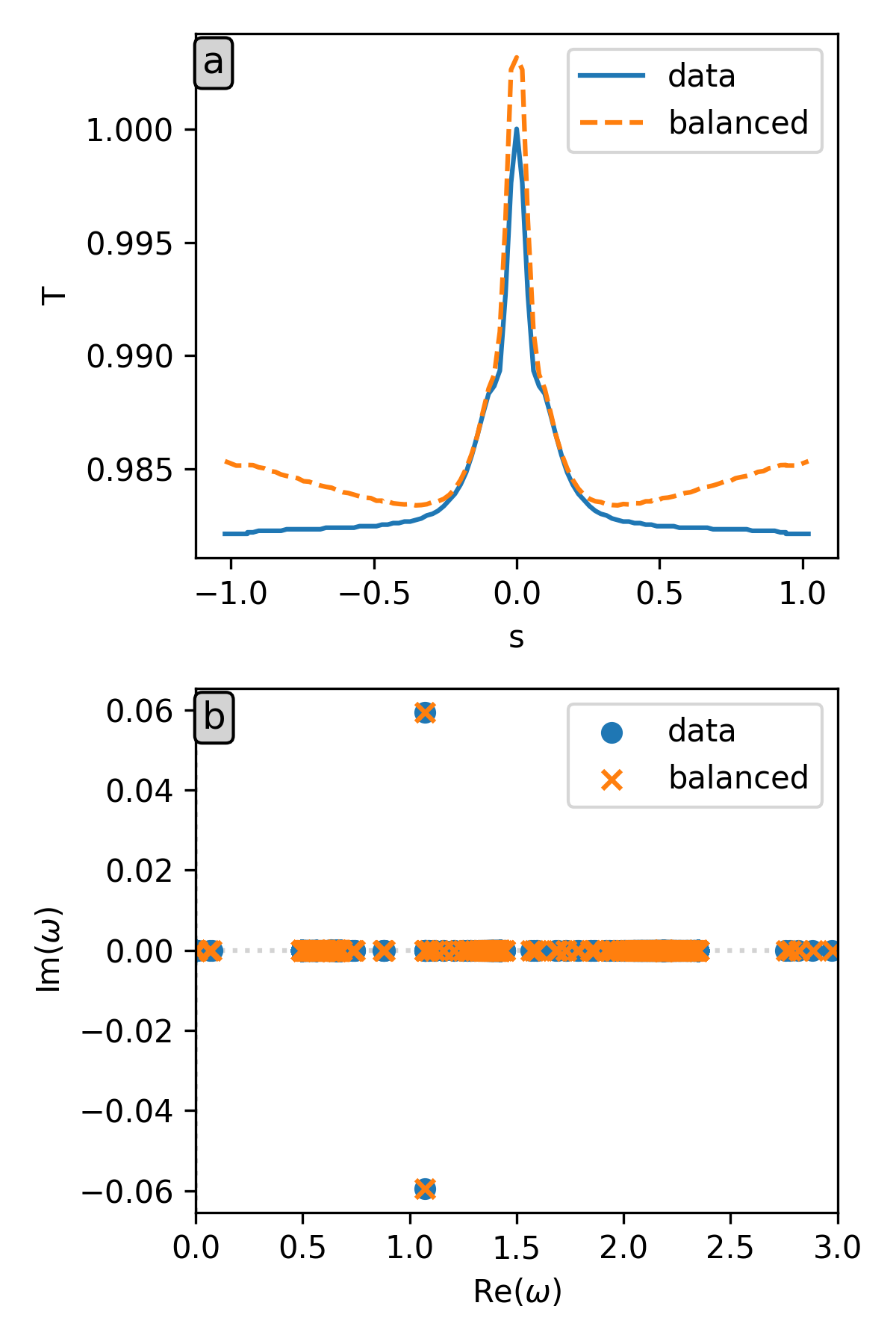}
    \caption{Comparison of (a) the B2T1.32 temperature profile and the force-balanced temperature profile for this case's density and magnetic field profiles; and (b) the corresponding spectra.}
    \label{fig:balanced}
\end{figure}
\end{appendix}

\end{document}